# Non-Hermitian two-dimensional photonic crystal flat lens

**Longfei Li, Jianlan Xie, and Jianjun Liu***

*Key Laboratory for Micro/Nano Optoelectronic Devices of Ministry of Education & Hunan Provincial Key Laboratory of Low-Dimensional Structural Physics and Devices, School of Physics and Electronics, Hunan University, Changsha 410082, China.*

*Corresponding author: jianjun.liu@hnu.edu.cn

**Abstract:** In this paper, a non-Hermitian two-dimensional photonic crystal flat lens is proposed. The negative refraction of the second band of photonic crystal is utilized to realize super-resolution imaging of the point source. Based on the principles of non-Hermitian systems, a negative imaginary part is introduced into the imaging frequency, in which case the imaging intensity and resolution are improved. The results indicate that the non-Hermitian system provides a new method to improve the imaging performance of the photonic crystal lens.

Photonic crystal (PC)[1,2] is a kind of artificial microstructure in which the dielectric constant changes periodically or quasi-periodically in space and it is constructed at the light wavelength scale. Due to the properties of photonic localization, photonic bandgap (PBG), and negative refraction, PC is widely utilized in the design of optical integrated devices, including optical fibers,[3-11] filters,[6,12] sensors,[8,11,13,14] couplers,[15,16] light-emitting diodes,[17] prisms,[18] lenses,[19-28] waveguides,[15,29,30] etc.

On account of characteristics of negative refraction, low loss and easy preparation, two-dimensional (2D) PC is extensively applied in lenses. In order to improve the imaging quality of 2D PC lenses, researchers has proposed trapezoidal PC lenses,[19] photonic quasicrystal (PQC) lenses,[20] and gradient PC lenses.[21,22,27,28] However, quasi- periodic structure and gradient make it difficult to prepare the lenses, resulting in the inability to promote them. Thus on the premise of ensuring the uniformity and periodicity of PC rods, it is impossible to improve the imaging performance of PC lenses greatly in the current Hermitian system. In recent years, there have been a lot of theoretical and applied researches[31-34] about non-Hermitian systems, such as electromagnetic diodes,[35] optical fibers,[36] couplers,[37] lasers,[38] waveguides,[39-41] topological insulators,[42] diffraction gratings,[43] metasurfaces,[44] resonators,[45] interfaces[46] and sensors[47]. The core idea is to bring spatially alternating gain and loss elements with the same intensity into the Hermitian system, ensuring a negative imaginary part introduced to the operating frequency, so as to make the properties of the integrated device advanced greatly. According to the definition of gain and loss,[48] when the imaginary part of the operating frequency is greater than zero (i.e., the positive imaginary part), it corresponds to loss. On the contrary, when the imaginary part is less than zero (i.e., the negative imaginary part), it corresponds to gain. If the non-Hermitian system can be applied in a 2D PC lens to introduce a negative imaginary part into the imaging frequency, it is expected to achieve optical gain, so that the imaging intensity and resolution of the flat lens are able to be improved without changing the uniformity and periodicity of PC rods, making it easier to put it into practical application.

In this paper, the imaging characteristics of a 2D PC flat lens in Hermitian and non-Hermitian systems are designed and analyzed based on the square lattice respectively. The results show that, compared with the Hermitian system, the imaging intensity is significantly enhanced and the imaging resolution is remarkably improved in the non-Hermitian system. Therefore, the non-Hermitian system is able to provide a new way to improve the imaging performance of the PC flat lens.

Before introducing the non-Hermitian system, the imaging characteristics of the Hermitian 2D square lattice PC flat lens are firstly analyzed, and its model is shown in Fig. 1. The Hermitian 2D square lattice PC flat lens takes air as background. The radius of the rods $r=0.23a$ and the refractive indexes of two different dielectric rod materials are $n_1=2$ and $n_2=4$.

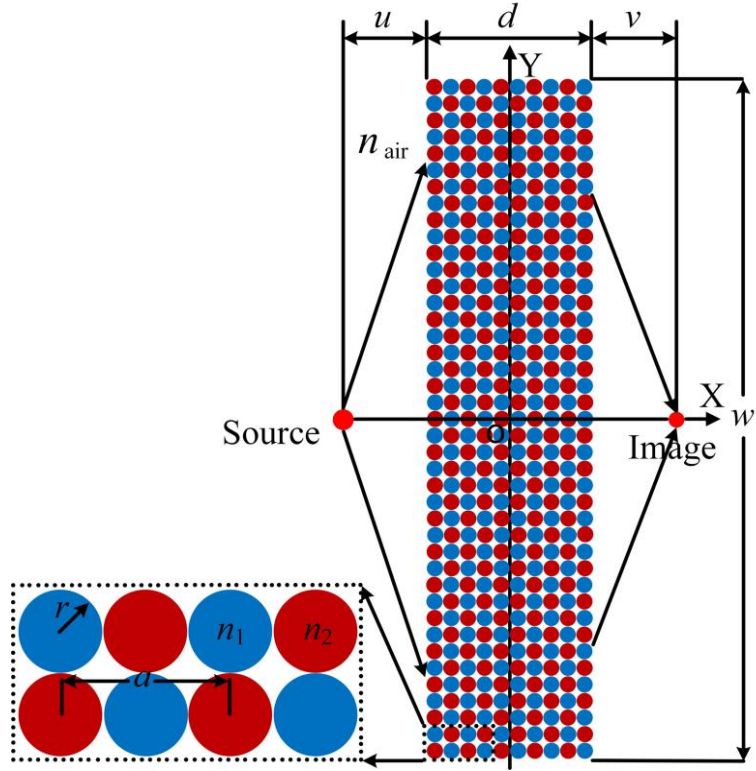

Fig. 1. The model of 2D square lattice PC flat lens.

The lattice constant $a=1\mu m$ remains unchanged throughout the lens, with the width of the lens $w=21a$, the thickness $d=5a$, the object distance $u=0.5d=2.5a$, and the image distance set as $v$. The finite element method is used to calculate and analyze the imaging characteristics of the lens in this paper.

When beam is incident from the air to the PC, the direction of the corresponding refracted beam is the same as the direction of the group velocity $v_g$. The group velocity $v_g$ and the wave vector $k$ satisfy the vector relationship[49]

$$v_g = \nabla_k \omega(\mathbf{k}) \qquad (1)$$

It can be found that the direction of the refracted beam is consistent with the frequency gradient of the equi-frequency surface (called equi-frequency line (EFL) in the case of 2D). Since the frequency gradient is related to the band structure of the PC, it is convenient to determine whether the 2D square lattice PC has a negative refraction effect by calculating the band structure. The specific theoretical analysis is shown in Fig. 2.

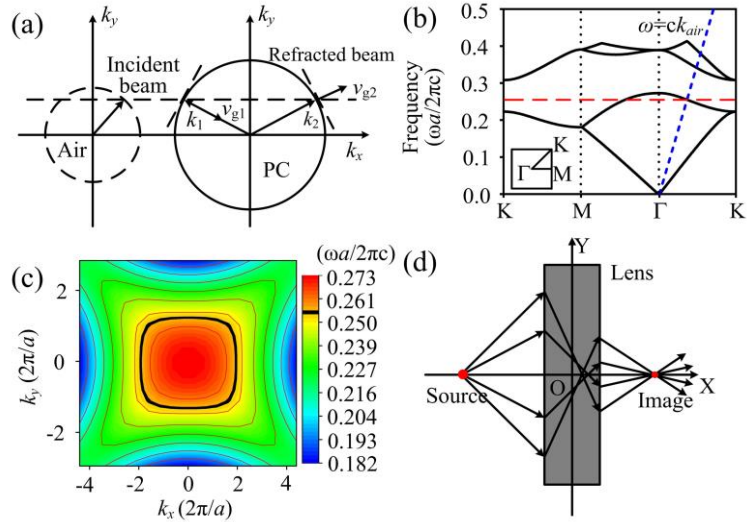

Fig. 2. (a) Direction of beam propagation in the PC is analyzed by the EFL method. (b) Partial band structure of Hermitian 2D square lattice PC. The blue dash line is the dispersion curve of air. The red dash line indicates the normalized frequency $f_1$=0.253 at the intersection of the second band and the dispersion curve of air. (c) EFL(s) of the second band, the black curve represents the EFL corresponding to the normalized frequency $f_1$. (d) Imaging model of the flat lens for the point source.

As can be seen from Fig. 2(a), the left circle represents the EFL when beam travels in air, and the right circle represents the EFL when beam travels in PC. When beam enters PC from the air, positive and negative refraction may occur. If the frequency gradually decreases radially from the center to the outward direction, the direction of the group velocity is the $v_{g1}$, opposite to the wave vector $k_1$. In this case, the refracted beam and incident beam are on the same side of the normal line as shown the horizontal dash line in Fig. 2(a), indicating that negative refraction occurs in the propagation process of beam. As shown in Fig. 2(b), there is a wide PBG between the second and the third bands. In the second band, the frequency decreases gradually from the center (Γ point) to the outward direction, so it is known that the band exists negative refraction. The EFL corresponding to the normalized frequency $f_1$ is convex, which is presented in Fig. 2(c), further confirming that the second band exists negative refraction and $f_1$ is the imaging frequency. The possible path of beam during propagation in the 2D PC with negative refraction is shown in Fig. 2(d).

The principle of why the non-Hermitian system can improve imaging performance is shown in Fig. 3.

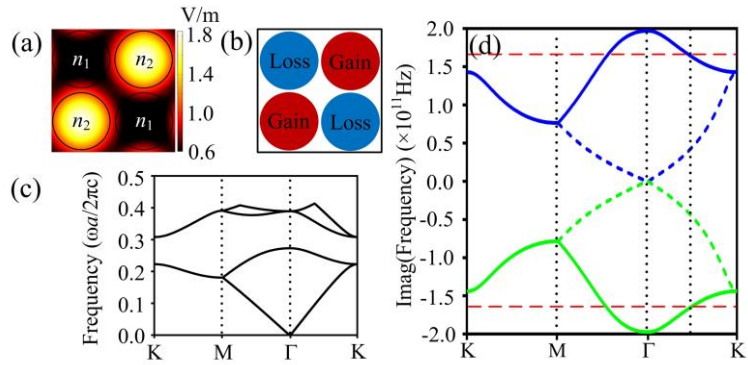

Fig. 3. (a) Mode field diagram corresponding to the imaging frequency $f_1$. (b) Two different dielectric rod materials are the optical gain and loss elements, respectively. (c) Partial band structure of non-Hermitian system with $β=0.01$; (d) The green dash and solid lines are the first and second bands of the imaginary band structure of the non-Hermitian system with $β=0.01$. The red dash line intersecting the solid green line indicates the negative imaginary part of $f_1$ (-1.65×$10^{11}$Hz). The blue dash and solid lines are the first and second bands of the imaginary band structure of the non-Hermitian system with $β$=-0.01. The red dash line intersecting the solid blue line indicates the positive imaginary part of $f_1$ ($1.65 \times 10^{11}$Hz).

    According to the mode field diagram corresponding to the imaging frequency $f_1$ which is shown in Fig. 3(a), most beam propagates through the dielectric rods with higher refractive index during the process of imaging. Since the ratios of two different dielectric rods are the same, if the gain element can be added on the main path of propagation (the dielectric rods with higher refractive index) with the loss element of the same intensity added on the secondary path of propagation (the dielectric rods with lower refractive index), the optical gain can be achieved on the premise of no net gain of the system itself. Therefore, the non-Hermitian system is introduced into the 2D PC flat lens. For the refractive indexes of two different dielectric rods, the imaginary part $β$ is added with the real part retained. The refractive indexes of two different dielectric rods are $n_{Gain}=n_2-iβ$ and $n_{Loss}=n_1+iβ$, respectively. Therefore, the intensities of gain and loss are controlled by the value of $β$ and are guaranteed to be the same. As is shown in Fig. 3(b), when $β>0$, the dielectric rods with higher refractive index are the gain elements and the dielectric rods with lower refractive index are the loss elements, respectively. At this point, the gain is introduced on the main path of propagation, and the optical gain can be realized. Conversely, when $β<0$, the loss is introduced on the main path of propagation. In this case, loss occurs during the propagation. Partial band structure of the non-Hermitian 2D square lattice PC with $β=0.01$ is presented in Fig. 3(c), which is consistent with the band structure of the Hermitian system which is shown in Fig. 2(b). Therefore, the imaging of the point

source at $f_1$ can also be realized in the non-Hermitian system. Further, taking $β=±0.01$ as the example, the imaginary part of the band structure of the non-Hermitian system is calculated, as shown in Fig. 3(d). It can be found that the imaginary parts of the band structure are mirror symmetrical. When $β=0.01$, $imag(f_1)<0$, the negative imaginary part is introduced into the imaging frequency $f_1$. The corresponding wave vector $k$ is in the gain state so that the optical gain can be realized. On the contrary, when $β=-0.01$, $imag(f_1)>0$, the positive imaginary part is introduced into the imaging frequency $f_1$. The corresponding wave vector $k$ is in the loss state, and beam is lost during propagation. As a result, the correctness of the design of the non-Hermitian 2D PC lens is demonstrated by the analysis of Fig. 3.

As can be seen from the analysis of Fig. 3, the non-Hermitian 2D square lattice PC flat lens can realize the imaging of the point source in the second band. The imaging frequency range is $f∈[0.245,0.261]$. In view of $λ=a/f$ and $a=1\,μm$, corresponding imaging wavelength range can be obtained as $λ∈[3.831\,μm, 4.082\,μm]$. Take the wavelength $λ_1=3.953\,μm$ which is in corresponding to $f_1$ as the example. The imaging of the point source in the second band is calculated as $β$ changes. The variations of the imaging intensity and FWHM with different values of $β$ are presented in Fig. 4.

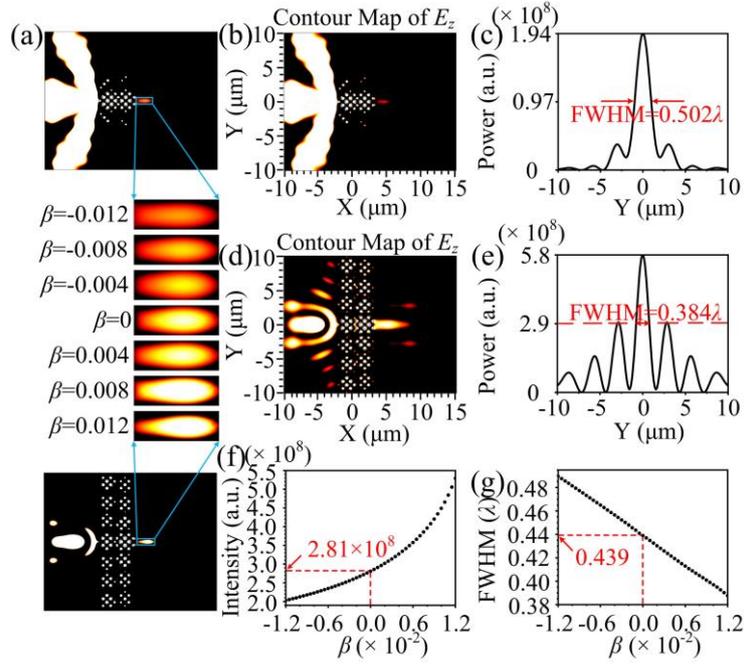

Fig. 4. (a) The comparison between image points with different values of $\beta$. (b)The imaging field of the point source when $\beta$=-0.013. (c) The field power (normalized) of imaging of the point source in the imaging plane when $\beta$=-0.013. (d) The imaging field of the point source when $\beta$=0.013. (e) The field power (normalized) of imaging of the point source in the imaging plane when $\beta$=0.013. (f) Relationship between the imaging intensity and $\beta$. (g) Relationship between FWHM and $\beta$.

As can be seen from Fig. 4(a), when $\beta$<0, optical loss occurs and the image point gradually darkens as $\beta$ decreases. On the contrary, when $\beta$>0, optical gain happens and the image point becomes brighter as $\beta$ increases gradually. As shown in Figs. 4(b) and 4(c), when $\beta$=-0.013, the optical loss makes the imaging intensity weak and it cannot break the diffraction limit (FWHM=0.502$\lambda$>0.5$\lambda$). According to Figs. 4(d) and 4(e), when $\beta$=0.013, the optical gain enhances the imaging intensity and breaks the diffraction limit (FWHM=0.384$\lambda$<0.5$\lambda$). However, in this case, the sideband energy is higher than half of the imaging intensity, which is not convenient for the detection of the image point and its practical application.[50] According to Figs. 4(f) and 4(g), when $\beta=0$, the image point is in the Hermitian system, its intensity is $2.81\times10^8$ (a.u.), and FWHM=$0.439\lambda<0.5\lambda$, which breaks the diffraction limit, that is, super-resolution imaging is achieved. In the non-Hermitian system, when $\beta\in[-0.012,0)$ (optical loss), as the $\beta$ decreases, the imaging intensity decreases monotonously, and the FWHM increases gradually, which means the image performance gets worsen gradually. When $\beta\in(0,0.012]$ (optical gain), as the $\beta$ increases, the imaging intensity increases monotonously and the increasing rate increases gradually (the slope of the curve increases gradually), and the FWHM decreases monotonously, so the image performance is improved gradually. When $\beta=0.012$, the imaging performance of the non-Hermitian 2D square lattice PC flat lens becomes the best. At the point, the imaging intensity can reach $5.29\times10^8$ (a.u.), which is 1.8826 times of the imaging intensity in the Hermitian system ($\beta=0$), and FWHM=$0.387\lambda<0.439\lambda$ ($\beta=0$). Compared with the Hermitian system, the resolution is improved greatly. Therefore, the introduction of the non-Hermitian system can improve the imaging performance of the 2D PC flat lens significantly. During the process of propagation, the intensity of the evanescent wave is enhanced. Compared with the Hermitian system, the sub-wavelength information is more retained. Therefore, when the electromagnetic wave propagates to the imaging plane, the information about the phase and amplitude of the point source are both better restored, which can greatly improve the resolution.[51]

The non-Hermitian 2D PC flat lens proposed in this paper can realize the super-resolution imaging of the point source. Compared with the Hermitian system, the imaging intensity is enhanced greatly and the resolution is improved obviously. The imaging performance of the 2D PC flat lens is improved under the premise of ensuring the uniformity and periodicity of PC rods.

**Funding.** National Natural Science Foundation of China (61405058); Natural Science Foundation of Hunan Province (2017JJ2048); Fundamental Research Funds for the Central Universities (531118040112); Student's Platform for Innovation and Entrepreneurship Training Program.

**Acknowledge.** The authors acknowledge Prof. J. Q. Liu for software sponsorship and the help of Hexiang Zhao and Yuan Cen.